\documentclass[a4paper,fleqn,usenatbib]{mnras}

\usepackage[T1]{fontenc}
\usepackage{ae,aecompl}
\usepackage{float}
\usepackage{graphicx}   
\usepackage{color}
\usepackage{amssymb}
\usepackage{amsmath}

\title[BW analysis of Blazhko stars in M3]{The first Baade-Wesselink analysis of Blazhko RR Lyrae stars: discrepancies between photometrically and spectroscopically determined radius variations}
\author[J. Jurcsik et al.]{J. Jurcsik$^{1}$\thanks{E-mail: jurcsik@konkoly.hu}, G. Hajdu$^{2,3}$
\\
$^{1}$Konkoly Observatory of the Hungarian Academy of Sciences, H--1525 Budapest PO Box 67, Hungary\\
$^{2}$Instituto de Astrof\'isica, Pontificia Universidad Cat\'olica de Chile, Av. Vicu\~na Mackenna 4860, 782-0436 Macul, Santiago, Chile\\
$^{3}$Instituto Milenio de Astrof\'isica, Santiago, Chile\\
}
\begin{document}
\date{Accepted 2017 ..... Received 2017 ..., in original form }

\pagerange{\pageref{firstpage}--\pageref{lastpage}} \pubyear{2017}

\maketitle
\label{firstpage}
\begin{abstract}
The simultaneous photometric and spectroscopic observations of the RR Lyrae variables in the globular cluster, M3, published in \citet[][Paper I]{data} made it possible to perform Baade-Wesselink (BW) analysis  of a large sample of Blazhko stars for the first time. The BW distances of Blazhko stars turned out to be unreliable, as significantly different distances were obtained for the stars of the Blazhko sample and also for the same star in different modulation phases. Even the results of small modulation-amplitude Blazhko stars may be doubtful. This result warns that the application of the BW method to Blazhko stars is not trustworthy.  

Keeping the distance fixed for each Blazhko star in each modulation phase, a significant difference between the spectroscopic and the photometric radius  ($R_{\textrm{sp}}$, $R_{\textrm{ph}}$) variations is detected. The phase and amplitude variations of $R_{\textrm{sp}}$ follow the changes of the light curve during the Blazhko cycle but the $R_{\textrm{ph}}$ curve seems to be not (or only marginally) affected by the modulation. The asynchronous behaviour of $R_{\textrm{sp}}$ and $R_{\textrm{ph}}$ supports the interpretation of the Blazhko effect as a depth-dependent phenomenon, as the spectroscopic radius variation reflects the radial displacement of the line-forming region high in the atmosphere, while the photospheric radius variation is derived from the information of the observed visual-band light emitted mostly by the lower photosphere.  The stability of  $R_{\textrm{ph}}$ may be interpreted as a strong argument against the non-radial-mode explanation of the Blazhko phenomenon.
\end{abstract}

\begin{keywords}
stars: horizontal branch --
stars: oscillations (including pulsations) --
stars: variables: RR Lyrae --
Galaxy: globular clusters: individual: M3 --
techniques: photometric --
techniques: radial velocities
\end{keywords}

\section{Introduction}

The main goal of the Baade-Wesselink (BW) method is to determine the distance and the radius of radial pulsators using photometric and radial-velocity (RV) data. The basic concept  is to measure the line-of-sight radial displacement of the stellar surface via spectroscopic tools, and the simultaneous angular-diameter variation of the stellar disk using photometric (recently also  interferometric) methods.  The distance and the mean radius value of the star are then determined by fitting the amplitudes of the two radii variations via appropriate scaling. The BW method, and its variants were successfully applied to several radial-mode pulsators, from the pre-/post- and main sequence  $\delta$Sct stars  (e.g. \citealt{bm}, \citealt{hae}) to the horizontal-branch giant RR~Lyrae (RRL) variables (without completeness, e.g. \citealt{jcsl,f94,ccf,lj90} and \citealt{sk93}) and to the super-giant Cepheids  (as a summary of the BW distances of Cepheids read e.g. \citealt{dib} and \citealt{gron}).

The BW analysis relies on some basic assumptions, which may hold some uncertainties. The exact value of the projection factor, which converts radial velocity to pulsation-velocity variation is still under active investigation \citep{ker,nar}, and any inaccuracy of the colour-temperature and bolometric-correction transformations -- based on synthetic atmosphere-model results -- can modify the BW results. As the radius variations measured spectroscopically and photometrically reflects the radial motions at different depths, and differences between the motions of these layers may occur in dynamic atmospheres, depth-dependent phenomena (shocks, velocity gradients, etc.) may have a strong bias on the BW distances. The applicability of the method also requires that the pulsation has to be fully radial.

 Though some of the conditions of the applicability of the BW method listed above are not fully satisfied during the pulsation of RRL stars (e.g. the shocks propagating in the atmosphere are strongly depth dependent), and any inaccuracy of the $p$-factor and the synthetic model-atmosphere data biases the BW results, the error of the BW distances of RRL stars has been estimated to be  $3-10$ per cent for galactic field and globular-cluster RRL stars in previous studies \citep[e.g.][]{jcl,ccp,storm2}.

The light curves of about half of the fundamental-mode RRL stars are not stable \citep{kbs, k10}; amplitude and phase variations, i.e. the Blazhko effect is detected. Recently, Blazhko modulation of the light curves of Cepheids in the Galactic field \citep{blcep} and the Magellanic Clouds \citep{so15,s17} has also been discovered. Moreover, \cite{ander14,ander16}  showed that the small-amplitude modulations of the RV curves observed in some Cepheids may affect their BW distances, too. 

The applicability of the BW method on Blazhko RRL stars may be questioned until the physical mechanism inducing the phenomenon is explained. Besides this, the requirement of simultaneous spectroscopic and photometric observations is the other reason why no BW analysis of any Blazhko star has been published yet.
Although extensive spectroscopic observations of some Blazhko RRL stars have been published \citep{cg96,cp13}, they lack simultaneous multi-band photometry needed for the application of the BW method.

To perform the first BW analysis  of a large set of RRL star in a globular cluster and of Blazhko stars in different phases of the modulation, simultaneous photometric and spectroscopic observations of the variables of M3 were secured in 2012. The results of the photometric observations of overtone/double-mode variables  were published in \cite{overtone}. The photometric data of RRab stars, together with the RV measurements  and the results of the BW analysis of single-mode variables were presented in Paper I. 

 BW distances of 26 stable-light-curve RRL stars of M3 were determined in Paper I. 10.5 kpc mean distance of the cluster with 0.2 kpc formal error of the mean value was derived. The random, uncorrelated errors connected to the uncertainties of the magnitudes of the individual stars and the fitting process of the $R_{\textrm{sp}}$ and the $R_{\textrm{ph}}$ curves are significantly reduced if the sample comprises stars of the same distance, as for globular clusters. Indeed, the 0.2 kpc error does not exceed the real dispersion of the distances of the stars, as the tidal radius of M3 is about 0.1 kpc. As the global properties,  which influence the systematic errors of the BW distances (reddening, [Fe/H], zero point of the photometry) are thought to be known with relatively high precision for M3, the major sources of the  systematic biases  are connected to the uncertainties of the $p$-factor and the synthetic model-atmosphere data. \cite{jcl} estimated the effect of these factors to be 0.13 mag on the distance modulus, i.e. 6 per cent of the distance. Using the more recent model-atmosphere data \citep{kurucz} and an updated value for the $p$-factor \citep{n04},  the systematic errors of the derived distances might be somewhat smaller, however, only the Gaia parallaxes  will determine the  accuracy of the BW distances correctly.
The  data published in Paper I provide the first observations suitable to perform direct BW analysis of Blazhko stars in different phases of the modulation. The obtained results are very intriguing, and rise suspicion  concerning the applicability of the BW analysis for Blazhko stars.
 The aim of this paper is to document these issues, and not to determine the actual distance and radius values of the studied stars.

\section{Data, method and BW distances of Blazhko stars}\label{2}

\begin{figure}
\centering
\includegraphics[width=9.6cm]{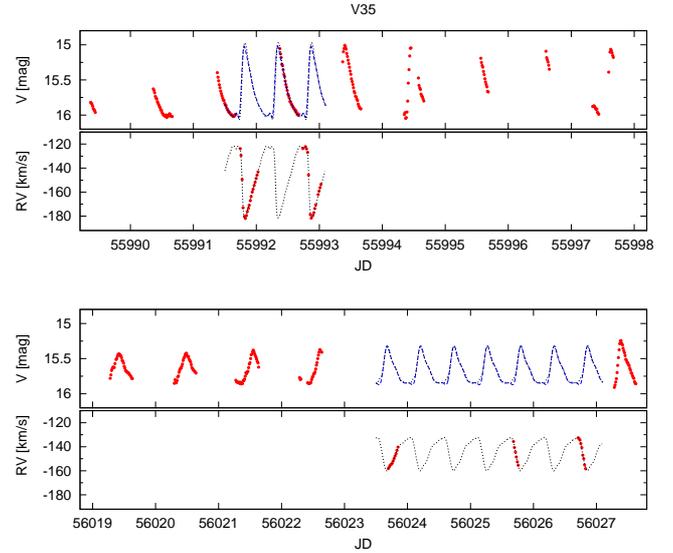}
\caption{The quasi-simultaneity of the photometric and spectroscopic observations is documented on the example of V035, a large modulation-amplitude Blazhko star with a 57-d modulation period. Nine-days-long segments of the $V$ light curve and the RV data are plotted; the single-mode Fourier solutions of the data, derived in two different ways for the light curve,  are drawn by dotted/dashed lines. The fits shown are used as the simultaneous $V$ and RV data in the analysis.   \label{v35}}
\end{figure}

\begin{figure*}
\centering
\includegraphics[width=18.1cm]{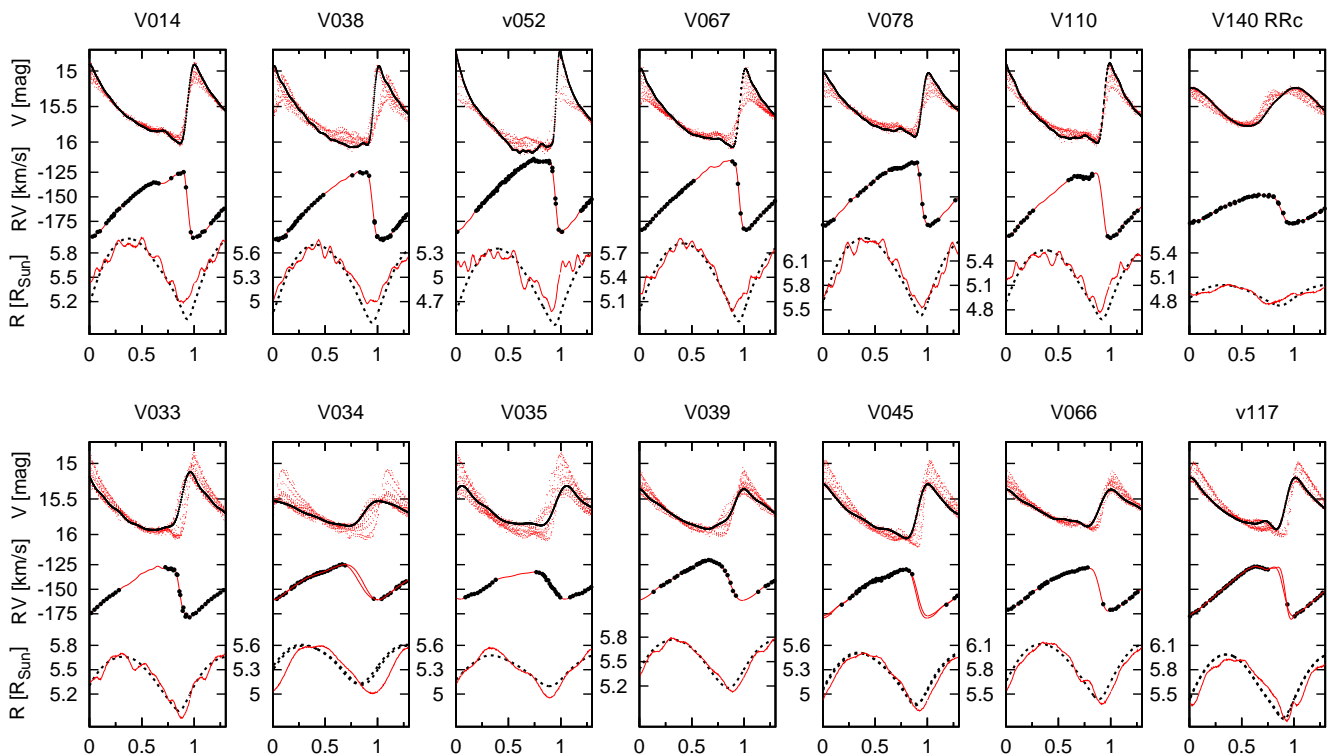}
\caption{Results of the  analysis of some Blazhko stars, whose RV data cover either maximum (top panels) or minimum (bottom panels)  Blazhko phases. The $V$ observations, with the synthetic light curve parallel with the RV data highlighted, the RV data and the fit, and the comparison of the $R_{\textrm{sp}}$ (dotted line) and  $R_{\textrm{ph}}$ (continuous line)  radius curves versus pulsation phase are shown. The analysis assumes $d=10.5$ kpc distance for each star. The amplitude of the $R_{\textrm{sp}}$ radius variation is larger than the amplitude of the  $R_{\textrm{ph}}$ curve at the large-amplitude phase of the modulation, and the difference is of the opposite sense at Blazhko minimum. Phase discrepancy between the $R_{\textrm{sp}}$ and $R_{\textrm{ph}}$ variations is also evident in some cases, especially if the phase modulation of the light curve is significant.}
\label{bw3x} 
\end{figure*}

The photometric and RV data of RRL stars in M3 were published in \cite{overtone} and in Paper I. The full description of the application of the BW method was also given in Paper I. 

Exclusively, the RV data derived from the Hectoechelle@MMT observations are used in the analysis of Blazhko stars, as they are condensed enough to provide accurate RV curves for separate Blazhko phases. 
The two runs of the Hectoechelle@MMT observations  cover the pulsation phases more or less completely at two epochs separated by $\sim30$ days for 14 stars showing light-curve modulation. The data are also suitable for the analysis of eight more Blazhko stars in one Blazhko phase. Most of these Blazhko stars are fundamental-mode RRab type variables, but RV data have been obtained for one of the overtone-mode RRc stars (V140) exhibiting Blazhko modulation, too. Altogether, we can study the results of the BW method in 36 discrete phases of the modulation using the data of 22 Blazhko stars.

The procedure of determining the  photometric data parallel with the RV observations for Blazhko stars  is twofold: single-mode Fourier fits of  $5-10$~d segments in the vicinity of the spectroscopic observations of either the direct photometric observations, or of the synthetic data generated according to the full light-curve solution are used. The differences between these synthetic data are, however, marginal, and the choice of using any of these solutions has no significant effect on the results. The photometric and spectroscopic data-coverage  in the vicinity of the two Hectoechelle@MMT runs is shown in Fig.~\ref{v35}, for a Blazhko star (V035) with a 57-d modulation period and with a strongly variable light curve, as an example. The dotted/dashed lines indicate the synthetic data used in the analysis. Two solutions for the synthetic $V$ light curve, derived as described above, are displayed.

The complete BW analysis carried out for  Blazhko stars has led to a  contradictory result. The derived distances cover  an unrealistically large, $7.7-14.5$ kpc range, which is significantly larger than the range of the BW distances obtained for stable-light-curve RRL stars. As a result of the BW analysis of 26 stable-light-curve RRL stars, 10.5 kpc mean value for the distance of M3 was derived in Paper I, with the individual distances falling in a narrower, $9.7 -11.4$ kpc distance range.  Although both the light and RV curves of Blazhko stars utilised in the analysis are less accurate than the mean light and RV curves of stable-light-curve variables, especially for variables with strongly and rapidly varying light curves, it cannot account for the large spread of the BW distances derived for Blazhko stars.

 The random errors of the individual distances of stable-light-curve RRL stars were dominated by the uncertainties arising from the selection of the phase interval omitted from the fitting process of the $R_{\textrm{sp}}$  and  $R_{\textrm{ph}}$ curves. Depending on the choice of the part of the $R$ curves to match, up to 0.5 kpc differences were obtained. This is the case for Blazhko stars, too, but the effect is as large as $1.0-1.5$ kpc for some stars. The error connected to the incomplete phase coverage of the  RV curve and the uncertainity of the light cuvre at the dates of the RV observations is tested using different possible solutions for the actual RV and light curves. The distances derived this way remain within $0.5-1.0$ kpc range, typically. The effect of the random errors of the photometric zero point of the individual stars on the distance estimates is supposed to be less than 0.3 kpc for both Blazhko and non-Blazhko RRL stars, as the accuracy of the zero points of the ISIS flux photometry depends on the accuracy of the magnitudes of the variables determined on the reference frames, which is $0.01-0.05$ mag depending on the crowding.
 These errors, added in quadrature, give an estimate of $1.1-1.8$ kpc for the accuracy of the individual distances of Blazhko stars. Consequently, the $\sim7$ kpc full range of the distance values obtained for Blazhko stars are as large as $4-6\,\sigma$ of the estimated random errors. 

Not only the scatter of the individual distances derived for Blazhko stars are larger than the errors would indicate,
but as large as  $3-5$ kpc  ($2-4\,\sigma$) differences between the two distances obtained for  Blazhko stars with RV data suitable for the analysis at two epochs are obtained.
Such a large discrepancy between the derived distances of the same star using different sets of the observations cannot have any physically plausible interpretation, if the basic assumptions of the BW analysis are valid in both modulation phases. 
Consequently, we have to conclude that the BW method fails to give reliable distances for Blazhko stars.

\begin{figure}[h]
\centering
\includegraphics[width=8.8cm]{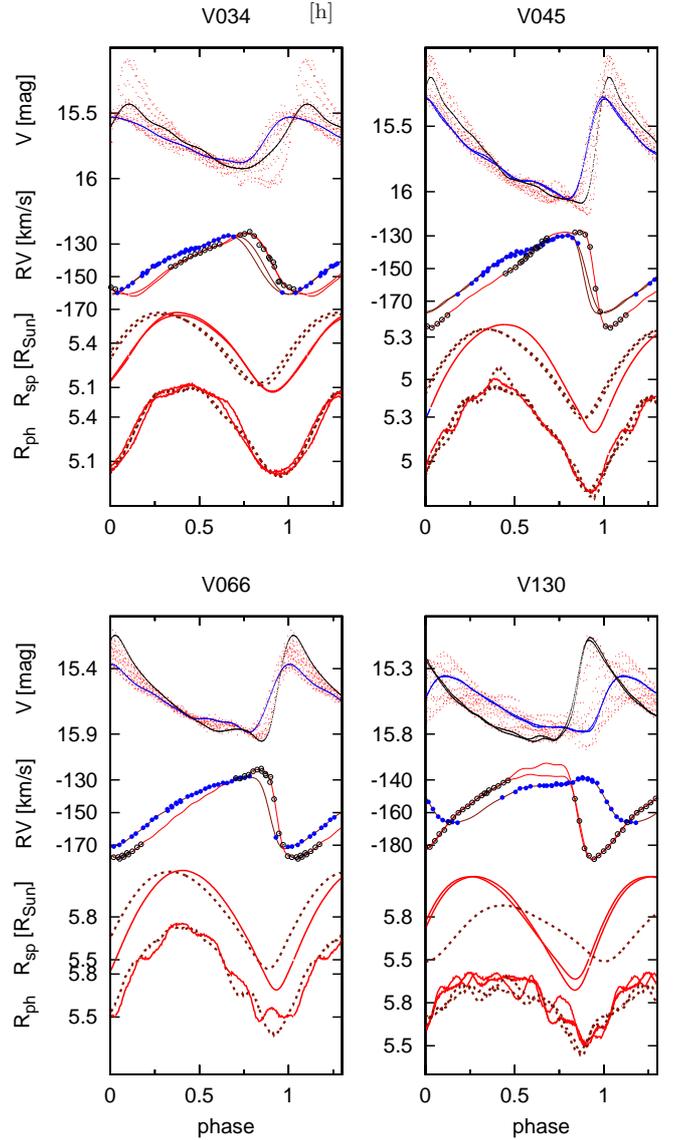}
\caption{Results of the  analysis of four large-modulation Blazhko RRab stars in two different phases of the modulation are plotted.  The full $V$ light curves, the light curves matching the two RV observation runs are highlighted, the RV observations and their fits, and the derived $R_{\textrm{sp}}$ and $R_{\textrm{ph}}$ variations are shown from top to bottom. The distance is fixed to $d=10.5$ kpc. The continuous and dotted lines denote the radius variations in the large- and the small-amplitude phases of the modulation. Some of the RV curves are scanty; different solutions to fill these gaps are shown for both the small- and the large-amplitude phases of V035, and for the small-amplitude phase of V045 and for the large-amplitude phase of V130. Two different representations of the photometric data  (as described in Sect.~\ref{2}) are considered for the small-amplitude phase of V045 and for both the small- and the large-amplitude phases of  V130, in order to indicate the effect on the resultant  $R_{\textrm{ph}}$ curves.
}
\label{bw4} 
\end{figure}

\begin{figure}[t]
\centering
\includegraphics[width=8.7 cm]{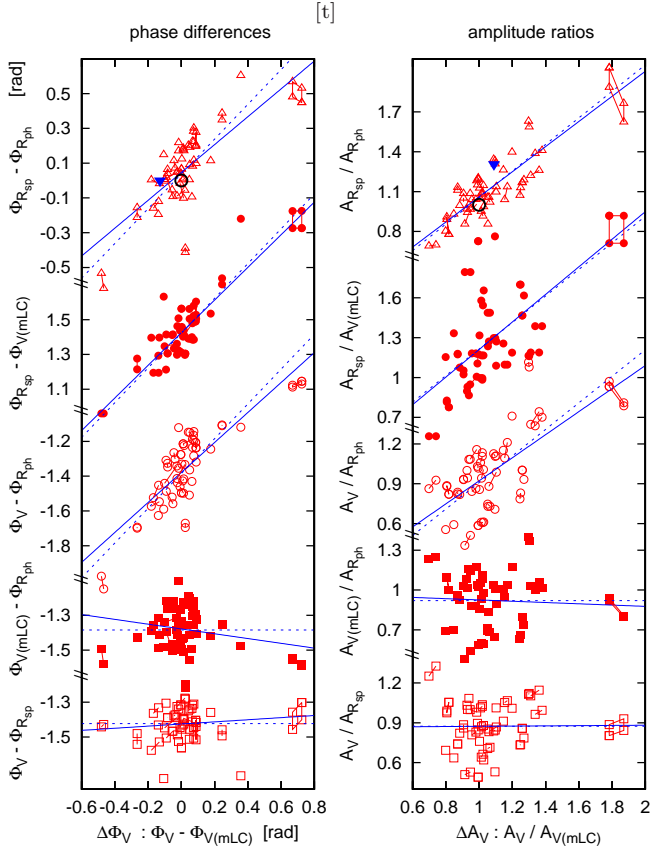}
\caption{Phase differences and amplitude ratios between different combinations of the $V$, $V_{\textrm{mLC}}$, $R_{\textrm{sp}}$ and $R_{\textrm{ph}}$ variations of Blazhko stars versus the phase difference (left-hand panel) and amplitude ratio (right-hand panel) of  $V$ and $V_{\textrm{mLC}}$. $V$ and $V_{\textrm{mLC}}$ denote the actual and the mean $V$ light curve, respectively. First-order Fourier amplitudes and phases are utilised. The first three plots from top (the relations between $R_{\textrm{sp}}$ and $R_{\textrm{ph}}$, between $R_{\textrm{sp}}$ and $V_{\textrm{mLC}}$ and between $V$ and $R_{\textrm{ph}}$) follow the amplitude and phase relations between $V$ and $V_{\textrm{mLC}}$.  The slope of the best-fit lines to the data (continuous lines) equals with 1.0 (dashed lines) within the uncertainty limits in these plots. The two bottom plots document the relations between $V_{\textrm{mLC}}$ and $R_{\textrm{ph}}$ and between $V$ and $R_{\textrm{sp}}$.  These parameters are independent from the phase and amplitude relations between $V$ and $V_{\textrm{mLC}}$, as the  steepness of the linear fits to these data are close to zero. The exact, one-to-one (three plots from the top) and zero-steepness relations (two plots from the bottom) are shown by dotted lines for comparison. The black circles denote the  0:0 and 1:1 positions in the topmost relations of the left- and right-hand panels, respectively. Different solutions for the RV fits of scanty data, and  for the light curves parallel with the RV observations  are used in the analysis for some of the stars. These multiple results are connected by lines. The differences between these data hardly exceed the symbols' size for most of the cases. 
The filled blue triangles denote the relations between $R_{\textrm{sp}}$ and $R_{\textrm{ph}}$ of V140, the only RRc star in the sample. The other phase and amplitude relations of V140 are not plotted, because they are  significantly different for overtone- and fundamental-mode variables.}
\label{bwbl} 
\end{figure}

\section{Discrepancies between $R_{\textrm{sp}}$ and $R_{\textrm{ph}}$  in Blazhko stars}\label{radii}

In order to find the reason of the failure of the BW method for Blazhko stars, we have reanalysed the data using a fixed, $d=10.5$ kpc distance, as determined for stable-light-curve variables of M3 (Paper I), for each modulated star in each modulation phase, with RV observations enough for the analysis. 

The results  are shown in  Fig.~\ref{bw3x} for some Blazhko stars, whose RV observations cover the maximum or the minimum phase of the modulation. The $V$ light curves, with the synthetic fit  matching the dates of the RV observations set out, the RV data and the Fourier fit, and the radius variations determined from the spectroscopic ($R_{\textrm{sp}}$: dotted lines) and the photometric ($R_{\textrm{ph}}$: continuous lines) data as described in Paper I, phased with the pulsation period are shown for 14 Blazhko stars according to the  fixed-distance BW solution. The top- and bottom-line panels correspond to Blazhko-maximum and Blazhko-minimum data, respectively.

What stands out most, is that the amplitude of the $R_{\textrm{ph}}$ variation ($A_{R_{\textrm{ph}}}$) seems to be smaller  than the amplitude of the $R_{\textrm{sp}}$ variation  ($A_{R_{\textrm{sp}}}$) at the large-amplitude phase of the modulation and, on the contrary, $A_{R_{\textrm{ph}}}$  is larger than $A_{R_{\textrm{sp}}}$ at around Blazhko minimum. Although the minimum of the $R_{\textrm{ph}}$ variation is less deep than the minimum of the ${R_{\textrm{sp}}}$ curve for many stable RRL stars (this is why this part is usually omitted from the BW analysis of large-amplitude variables exhibiting strongly violent atmospheric dynamics in these pulsation phases),  the opposite sign difference that is detected in the small-amplitude phase of the modulation, is not observed in any of the stable RRab stars as documented in the right-hand panel of figure 8. in Paper I.

Phase shift between the $R_{\textrm{sp}}$ and $R_{\textrm{ph}}$ curves is also evident in some Blazhko stars (V034, V045, V066, V117 and V140); besides the amplitude modulation of the light curve the phase modulation  also has a large amplitude in these stars. 

The behaviour of the $R_{\textrm{sp}}$ and the $R_{\textrm{ph}}$ curves shown in Fig.~\ref{bw3x} suggests that, while $R_{\textrm{sp}}$ follows the amplitude and phase changes observed in the light and RV variations, the $R_{\textrm{ph}}$ curve remains relatively stable in both amplitude and phase. 

This possibility is  checked first by using  Blazhko stars that have suitable data for the analysis at two significantly different  phases of the modulation. Fig.~\ref{bw4} illustrates the results for four large modulation-amplitude Blazhko stars in different phases of the modulation. Although the RV curve is somewhat ambiguous because of the scanty data in one or both Blazhko phases for these stars, it  has  only a marginal effect on the results, as documented in the figure.  Different solutions to fill in the gaps of the RV curves are shown for V034, V045 and V130, but the differences between the resultant $R_{\textrm{sp}}$ curves are negligible.
The uncertainties of the $R_{\textrm{ph}}$ curves arise from the uncertainty of the synthetic light and colour curves simultaneous with the RV data. As described in Sect.~\ref{2} it was derived either according to the full-light-curve solution, or from the photometric observations obtained close in time to the RV measurements. $R_{\textrm{ph}}$ curves derived using both of these methods are shown for V045 and V130 in Fig.~\ref{bw4}. Again, the difference between these results is much smaller than the phase and/or amplitude difference between the simultaneous $R_{\textrm{sp}}$ and $R_{\textrm{ph}}$ curves. Consequently, the uncertainties involved in the method cannot account for the discrepant behaviour of the $R_{\textrm{sp}}$ and $R_{\textrm{ph}}$ curves.

Each of the four stars  shown in Fig.~\ref{bw4} exhibits large-amplitude phase and amplitude modulations. As supposed to be normal, both the RV and the $R_{\textrm{sp}}$ variations  follow the amplitude- and the phase-changes of the light-curve. In contrast, the $R_{\textrm{ph}}$ curves seem,  indeed, to remain unaffected by the modulation in all of the four stars shown as examples. Therefore, these results support the assumption that the shape of the $R_{\textrm{ph}}$ curve is not affected by the modulation.

Secondly, the amplitude and phase stability of the $R_{\textrm{ph}}$ curves of Blazhko stars has also been checked using the data of all the Blazhko stars observed spectroscopically. Phase differences and amplitude ratios of the different combinations of the actual ($V$)  and the mean light curves ($V_{\textrm{mLC}}$) and the $R_{\textrm{ph}}$ and $R_{\textrm{sp}}$ radius variations  are examined.  The amplitudes and phases of the first order of appropriate-order Fourier-series solutions are considered for all the parameters investigated. These are less sensitive to any defect of the $R_{\textrm{ph}}$ appearing in the $0.8-1.0$ pulsation-phase interval in the large-amplitude phase of the modulation than the values of the higher-order Fourier components or the total amplitude of the variations and the phase of the absolute minimum values. Moreover, the first-order components can be determined accurately even for the very sinusoidal shape $R$ curves, which higher-order Fourier components are uncertain. The  amplitude and phase of the $V_{\textrm{mLC}}$ are derived from the fit of the full $V$ light curve.

The phase differences and amplitude ratios between the parameter pairs, ($R_{\textrm{sp}};R_{\textrm{ph}}$),   ($R_{\textrm{sp}};V_{\textrm{mLC}}$), ($V;R_{\textrm{ph}}$), ($V_{\textrm{mLC}};R_{\textrm{ph}}$) and ($V;R_{\textrm{sp}}$), versus the phase and amplitude relations of the actual and the mean light curves for 36 different Blazhko phases of 22 Blazhko stars  are shown in  Fig.~\ref{bwbl}. The linear fits to the data are drawn in the plots by continuous lines. For comparison, dotted lines represent the corresponding linear fits with 1.0 and 0.0 steepness.

If the hypothesis that the changes of $R_{\textrm{sp}}$ follow the modulation of the light curve  but the Blazhko effect does not influence the variations of $R_{\textrm{ph}}$ is true,  then the phase differences and amplitude ratios between $R_{\textrm{sp}}$ and $R_{\textrm{ph}}$ should have to vary according to the phase differences and amplitude ratios between the actual and the mean light curves. Fig.~\ref{bwbl} documents that this is indeed what happens. The steepness of the fitted lines to the $R_{\textrm{sp}}$ and $R_{\textrm{ph}}$ phase differences and amplitude ratios versus $\Phi_V - \Phi_{V_{\textrm{mLC}}}$ and $A_V / A_{V_{\textrm{mLC}}}$ (the top row in the figure) equals to 1.0 within the limits of the uncertainty, and the same is true for the relations between $R_{\textrm{sp}}$ and  $V_{\textrm{mLC}}$  and  between the actual light curve and $R_{\textrm{ph}}$, as well.

The two bottom plots in the panels of Fig.~\ref{bwbl} document the same result from the opposite aspect; if the amplitudes and phases of $R_{\textrm{ph}}$ and $R_{\textrm{sp}}$ are locked to the amplitudes and phases of the mean and the actual light curves, respectively, then the relations between $R_{\textrm{ph}}$ and the mean  light curve and between $R_{\textrm{sp}}$ and the actual light curve has to remain constant, i.e. they are independent from the amplitude and phase relations between the actual and the mean light curves. As the steepness of the fitted lines equals to zero  within the error ranges, these relations also verify  our statement.

The 0:0 and 1:1 coordinates, marked by black circles in the first row of the left- and right-hand panels of Fig. ~\ref{bwbl}, correspond to the positions where the phases and the amplitudes of both the actual and the mean light curves,  and the $R_{\textrm{ph}}$ and $R_{\textrm{sp}}$ curves are equal. These positions are very close to the fitted lines; according to the best linear fits, the phase difference and amplitude ratio between the actual and the mean light curves are $-0.05$ rad and 0.94 where the amplitudes and the phases of the $R_{\textrm{ph}}$ and $R_{\textrm{sp}}$ curves are the same.
Correspondingly, the phases of $R_{\textrm{ph}}$ and $R_{\textrm{sp}}$ coincide the best at the Blazhko phase when the first-order Fourier phase of the actual light curve is the same or it is slightly smaller than the phase of the mean light curve, i.e. the actual light curve is at the same phase or is shifted to slightly larger phase values than the mean light curve on a phased light-curve plot. Similarly, the amplitudes of the $R_{\textrm{ph}}$ and $R_{\textrm{sp}}$ radius curves are matching the best when the first-order Fourier amplitude of the actual light curve is close to or is a bit smaller than the amplitude of the mean light curve.

However, this result does not necessarily imply that the pulsation of Blazhko stars  would be normal when the  actual light curve is close in amplitude and phase to the mean light-curve, as the amplitudes of the mean light curves of most of the Blazhko stars are significantly smaller than the amplitudes of similar-period non-Blazhko stars (see figure 1. in Paper I).

Although there is no doubt that  the amplitude ratios and phase differences shown is Fig.~\ref{bwbl} confirm  that the variations of  $R_{\textrm{sp}}$ reflects the variations of the light curve, while the amplitude and phase of $R_{\textrm{ph}}$ remain relatively stable during the Blazhko cycle, the scatter of the plots shown in Fig.~\ref{bwbl} is notable; it is substantially larger than the uncertainties would indicate. Therefore, the outlined regularity has to be taken as a tendency rather than a rule.

\section{BW results for small modulation-amplitude Blazhko stars}\label{small}

\begin{figure*}
\centering
\includegraphics[width=17.9cm]{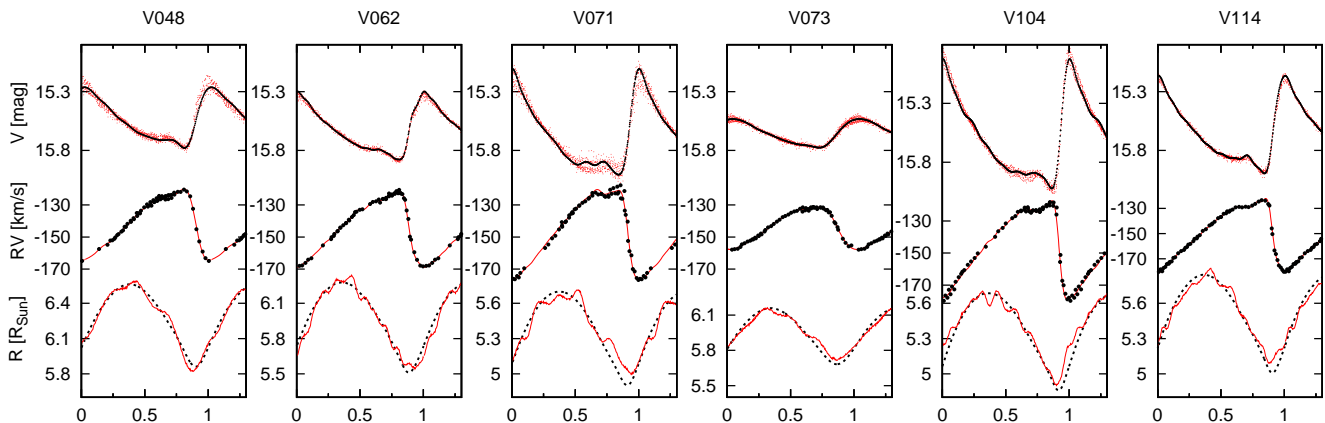}
\caption{Results of the analysis of the six small modulation-amplitude Blazhko stars using all the RV data and photometric data covering the epochs of the RV observations. The distance is fixed to 10.5 kpc. }
\label{bw6} 
\end{figure*}

\begin{table} 
\begin{center} 
\caption{Results of the BW analysis of six small modulation-amplitude Blazhko RRL stars. \label{satab}} 
\begin{tabular}{ll@{\hspace{2mm}}l@{\hspace{2mm}}rrr}
\hline  
Var.&P$_{\mathrm{puls}}$ [d]& P$_{\mathrm{Bl}}$ [d]&\multicolumn{3}{c}{distance rms [kpc] } \\ 
          &        &      &  \multicolumn{1}{c}{$a^*$}& \multicolumn{1}{c}{$b^{**}$}&  \multicolumn{1}{c}{all}\\   
\hline
V048 OoI? & 0.62783& 150   &  8.4 0.2 &          &  9.5 0.2\\
V062 OoI  & 0.65240& 280   &          &          & 10.2 0.4\\
V071 OoI  & 0.54905& \,46.7& 10.6 1.7 & 13.1 1.0 & 11.3 1.4\\
V073 OoI  & 0.67350& \,52.5& 13.7 0.3 &  8.6 0.3 & 12.4 0.2\\
V104 OoII & 0.56993& 108   & 11.0 0.4 &  9.5 0.4 & 10.8 0.4\\
V114 OoI  & 0.59773& \,54.4&          &          & 10.6 0.6\\
\hline
\multicolumn{6}{l}{$^*$ first run of the spectroscopic observations.}\\
\multicolumn{6}{l}{$^{**}$ second run of the spectroscopic observations.}\\
\end{tabular} 
\end{center}
\end{table}
\begin{figure}
\centering
\includegraphics[width=9.cm]{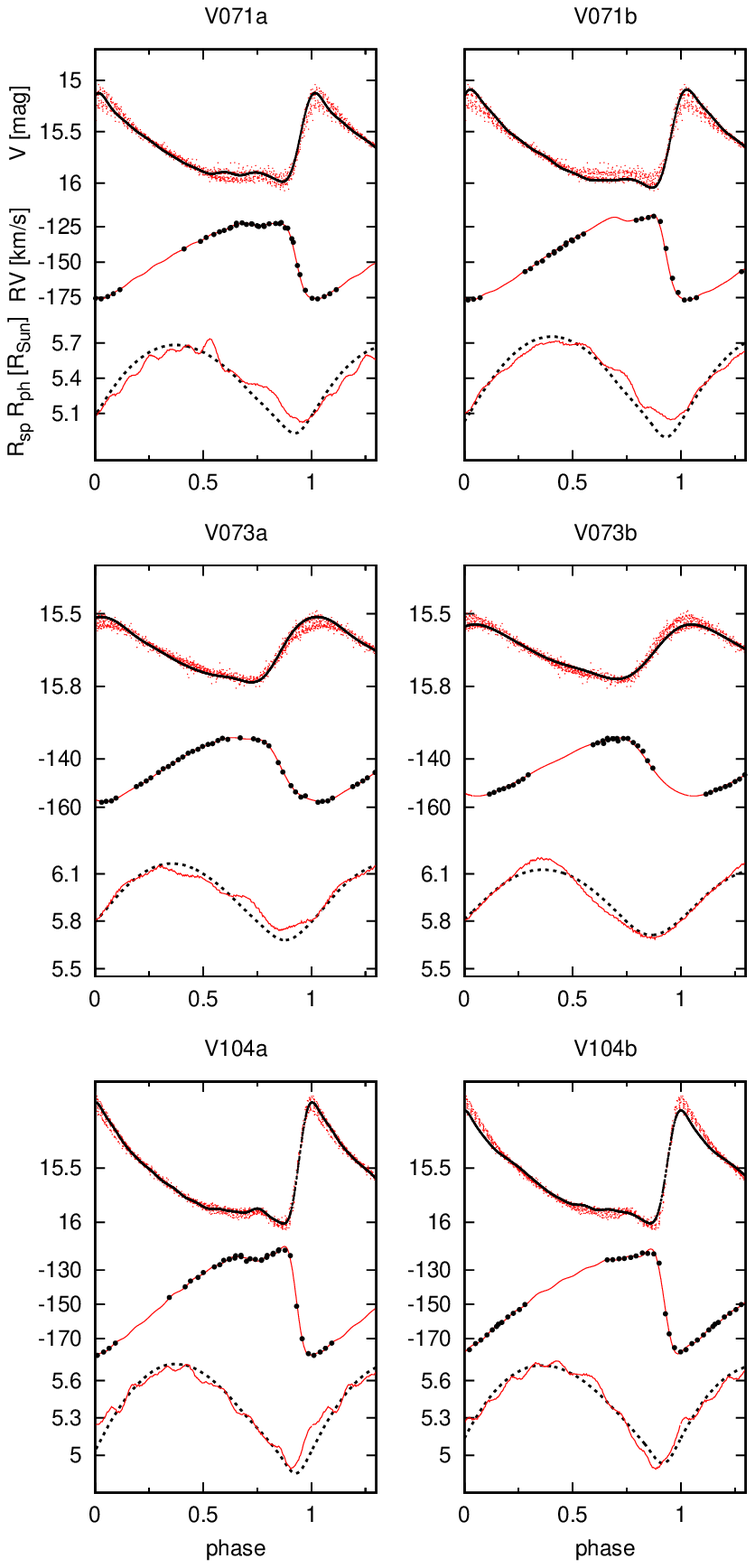}
\caption{Results of the analysis of three small modulation-amplitude Blazhko stars in two different phases of the modulation; $a$ denotes the first, $b$ the second run of the observations. The distance is fixed to 10.5 kpc. Note that the amplitude of $R_{\mathrm {ph}}$ is smaller than the amplitude of $R_{\mathrm {sp}}$ in the largest amplitude phase of the modulation (V071b and V073a), and $R_{\mathrm {ph}}$ has somewhat larger amplitude than $R_{\mathrm {sp}}$ in the smallest amplitude phase (V073b and V104b) even in these marginally modulated stars, similarly to the detected behaviour of large modulation-amplitude stars.}
\label{bwsa} 
\end{figure}

We have  investigated the small modulation-amplitude variables separately, in order to decide how reliable the BW distances are if the modulation  affects the light-curve shape only marginally.

RV data suitable for the BW analysis of six  small modulation-amplitude Blazhko stars were  published in Paper I. The same results as  shown for the Blazhko maximum and minimum phases of large modulation-amplitudes RRL stars in Fig.~\ref{bw4}, are shown  for the six small modulation-amplitude Blazhko stars in Fig.~\ref{bw6}. The distance is fixed to 10.5 kpc here, too.  The BW results shown in Fig.~\ref{bw6} were obtained by utilising the  combined set of the RV data of the two observation runs, and the corresponding synthetic light curves were determined form the segments of the photometric time-series close in time to the spectroscopic measurements. Although the matching of the $R_{\textrm{sp}}$ and $R_{\textrm{ph}}$ curves seems to be satisfactory for most of these stars, some discrepancy between the amplitudes seems to be present, e.g. for V071 and V073.

The results of the BW analysis of the small modulation-amplitude stars, with the distance as a free parameter, are summarised in Table~\ref{satab}. The pulsation and Blazhko periods and the BW distances derived from the data of the first and the second run of the spectroscopic observations (if any or both can be determined separately), and from the complete RV data set are given in the columns. As the photometric zero-point errors of the stars are the same in each phases of the modulation, the errors of the distances are estimated by fitting different parts of the $R$ curves. Table~\ref{satab} lists the mean values of the distances and their rms values derived for the data sets from 16 different solutions omitting $0.1-0.4$ phase intervals at around the minimum-to-maximum phases from the fitting process.

Comparing the results given in Table~\ref{satab} with the $10.5\pm0.2$ kpc mean value of the distances derived for stable RRL stars, we find that about half of the BW distances  of the small modulation-amplitude Blazhko stars determined for a given phase of the modulation  differ by $2-10\,\sigma$, $2-3$ kpc, from the mean, and for V048 and V073 the distances derived from the mean light curves and the complete RV data differ  by  $5-10\,\sigma$ from $10.5$ kpc, as well. 

The mean value of the distances derived using all the RV data of the small modulation-amplitude stars is 10.8 kpc with 0.4  kpc formal error. Omitting V048, the mean is $11.1\pm 0.4$ kpc (see the remark on V048 at the end of this section). This is $0.6$ kpc larger than the mean value of the distances derived for stable RRL stars, and its error is also larger than obtained for the non-Blazhko stars. However, the $1.5\,\sigma$ difference  between the mean values of the small-modulation-amplitude Blazhko and the non-Blazhko RRL stars is not statistically significant.

The analysis of the small modulation-amplitude Blazhko stars with good phase-coverage RV curves obtained for two epochs indicate the same systematic bias of the results as obtained for the large modulation-amplitude stars (see Fig.~\ref{bwsa}). Namely, the amplitude of  $R_{\textrm{sp}}$ is smaller than the amplitude of $R_{\textrm{ph}}$  at the small-amplitude phase of the modulation (V073b and V104b), and  the opposite is true at the large-amplitude phase (V071b and V073a).

Consequently,  the BW analysis may result in erroneous distance estimate even for very small modulation-amplitude Blazhko stars both if  data of a given Blazhko phase and if the mean RV and light-curve  are considered in the analysis.

{\it Remark on V048:} The photometric data of V048 is  contradictory, according to the mean $BVI$ magnitudes the star is a luminous, OoII variable, however, its amplitude is hardly larger than the amplitudes of similar-period OoI RRab stars in M3, even at the largest-amplitude phase of the modulation. The  amplitudes and mean magnitudes of the star according to previous photometric studies \citep{ca05,be06} indicated similar ambiguities. One possibility to resolve this inconsistency is to assume that the star is an OoI variable but, to explain the 0.1 mag too bright magnitudes, it is supposed to be closer by $\sim0.5$ kpc to us than the cluster itself. However, the $-141.4$\,km\,s$^{-1}$ mean RV of V048 is only 5.5\,km\,s$^{-1}$ larger than the  cluster's average RV, i.e. the difference is significantly smaller than the escape velocity. Therefore, it is unlikely that this star would indeed be an `escaper'. Most probably, an unresolved companion contaminates the photometry, making the star brighter and of smaller amplitude than normal. If this is the case, the BW analysis of V048 is unreliable independently from the Blazhko effect of the star.

\section{Discussion and conclusions}

We have shown in Sects.~\ref{radii} and ~\ref{small} that the discrepant behaviour of the $R_{\textrm{ph}}$ and the $R_{\textrm{sp}}$ radius variations leads to the failure of the application of the BW method for Blazhko stars.
As the BW method determines the distance via appropriate fitting of the amplitude of the $R_{\textrm{ph}}$ radius curve (angular radius variation) to the amplitude of the $R_{\textrm{sp}}$ curve (line-of-sight radius variation), the divergent behaviour of  their amplitudes and phases is the reason of the inadequacy of the BW method for Blazhko stars. Applying the BW method, the distance will be overestimated  if $R_{\textrm{ph}}$ has a smaller amplitude than $R_{\textrm{sp}}$ (around the large-amplitude phase of the modulation) and it will be underestimated  if $R_{\textrm{ph}}$ has a larger amplitude than $R_{\textrm{sp}}$ (around the small-amplitude phase of the modulation).

What is the origin of the discrepant behaviour of the $R_{\textrm{sp}}$ and the $R_{\textrm{ph}}$ curves  of Blazhko stars? 
A natural explanation would be that the pulsation of Blazhko stars is not fully radial; non-radial components also act as proposed e.g. in \cite{dm}.
However, the relative stability of the amplitudes and phases of the $R_{\textrm{ph}}$ curves compared to the variations of $R_{\textrm{sp}}$ should imply that the orientation of the non-radial oscillation is the same for each star and at each Blazhko phase, that is obviously far from being realistic.

Another possibility is that we see a depth dependent phenomenon.  \citet{sch} was the first who proposed to explain the observed line doubling and H$\alpha$ emission of pulsating variables by atmospheric shocks.
Following the pioneering hydrodynamical study \citep{h72} of  modelling the shock propagation, \citet{fo92} showed that two shocks are successively generated during one period in RR Lyrae, and that the main shock reaches its maximum amplitude very high in the atmosphere. Based on observations and model predictions, more details on the complexity of the dynamics of the atmosphere of RR~Lyrae stars, and on the role of the different shocks  in explaining the Blazhko effect were discussed  already in several papers \citep{mat,cg98,cvg08,gi13,cp13,gf14}. 

The detected strong correlation between the strength of the amplitude and phase modulations  and the amplitude and phase differences between the radial motions of the different depth layers reflected by the changes of $R_{\textrm{sp}}$ and $R_{\textrm{ph}}$ as shown in Fig.~\ref{bwbl}, indicates also that the atmospheric shocks responsible for the desynchronisation of the different depth layers have indeed an important role in the explanation of the Blazhko phenomenon.

Looking at the details, \cite{c14}  postulated that the Blazhko effect is generated by the interaction of a multi-shock structure with an out-flowing wind in the corona. The detection of  HeII emission at the large-amplitude phase of the modulation  \citep{p11,gfl13} showed that the dynamics of the higher atmosphere of Blazhko stars can be even more violent than the dynamics of stable RRab stars. 
\citet{gi13} proposed that the intensity of the main shock, producing the bump preceding light minima, varies during the Blazhko cycle. According to his description,  after reaching a critical strength of the main shock, the photospheric motions become desynchronised at around Blazhko maximum. A variant of this idea was suggested in \citet{cp13}. Although the phenomena delineated in these papers differ in details, both description predict that the atmospheric motions become synchronised at Blazhko minimum and are desynchronised at Blazhko maximum in order to explain the observed Van Hoof effect and the doubling/broadening of metallic lines at Blazhko maximum \citep{cp13}.

The observed variations of the $R_{\textrm{sp}}$ and the $R_{\textrm{ph}}$ curves also indicate  that the motions of the atmosphere are strongly desynchronised. However, it seems that the radial displacement of line-forming regions of the upper part of the atmosphere is synchronised with the radial displacement of the lower, photospheric layers, the measured light and colour information originate from, during the modulation phase when the amplitude and the phase of the light variation is close to (or is the same as) the amplitude and phase of the mean light curve. Significant desynchronisation between the different atmospheric layers occur both at the large- and the small-amplitude phases of the modulation.  

Despite of the large variety of the amplitude and phase modulations of the visual-band light curves of Blazhko stars, there is an indication that in the $K$ band, only marginal if any modulation is detected \citep{n15}. Hence, the pulsation seems to be stable at the depth of the $K$ band radiation.

Summarising all these results, we have information on the modulation of the pulsation from three different depths. First, the RV of the metallic lines reflects the motion of the topmost, line-forming region. Then, we can analyse the variations of the integrated visual-band light, radiated by the photosphere, and finally the K band variations hold information on the deepest regions of the photosphere, as the K-ban radiation comes from a larger depth than the visual-band light because of the near-IR dip in the opacities.

The RV variation of the metallic lines shows that the Blazhko effect manifests in this region as strongly varying  radius changes of these layers during the modulation cycle. 
The observed modulation of the visual-band light and colour curves corresponds to changes in the pulsation temperature and luminosity variations of the photosphere during the Blazhko period. At the same time, only marginal if any modulation of the $R_{\textrm{ph}}$ curves -- derived from the same visual-band photometric information -- is detected. It means that at the depth of the optical-band  radiation no or marginal modulation characterises the pulsation radius variation. Finally, at an even larger depth, the modulation seems to completely disappear, as the K-band data tends to indicate. 

What is the most surprising, is that not only the amplitude modulation disappears at a larger depth, but the phase modulation, too. 
Although the phase shift between the radial-velocity curves of V130 is as large as 1.3 radian ($0.2P_{\mathrm{puls}}$), no phase shift between the $R_{\textrm{ph}}$ curves is evident as shown in Fig.~\ref{bw4}. The phase modulation of the Blazhko phenomenon was interpreted as changes in the pulsation period e.g. by \cite{st,mw} and \cite{c14}. However, the recent result seems to contradict this possibility, as 
at the engine of the pulsation, at the partial ionisation zones, the period looks to be stable as indicated by the stability of the $R_{\textrm{ph}}$ and the $K$-band light curves. The modulation affects (both in phase and in amplitude) the temperature and luminosity variations during the pulsation of the inner part of the photosphere first, then modify the radial displacement of the outer part of the envelope as well.

As a summary, the first attempt to perform BW analysis of Blazhko stars in different phases of the modulation led to the conclusion that the modulation is a strongly depth-dependent phenomenon, that influences the pulsation of the up-most regions of the atmosphere only. This result supports the interpretation of the phenomenon  by the interaction of different shocks instead of the non-radial or the resonant-mode \citep{kb} explanations, as the detected depth dependency of the modulation properties seems to contradict the latter ideas. It is hard to reconcile  non-linear mode interactions with the strictly repeating nature of expansion and contraction of the deeper layers of the atmospheres of Blazhko RRL stars, as inferred from the present study, and from the reduced/missing nature of the modulation in the $K$-band. 

However, to prove that, indeed, shock interaction is  responsible for the modulation of the pulsation, 3-dimension modelling of the propagation of the different shocks in the atmosphere of RRL stars would be needed. It has to be noted as well that there are some properties of the Blazhko phenomenon, e.g. the relative stability and regularity of the modulation in some cases, and the lack of any connection between the physical properties of the stars and the occurrence of the modulation, which are hard to explain if the modulation is induced by shock interaction. Many details need to be investigated further in order to understand how the shock model of the modulation works. It would be desirable to observe at least one Blazhko RRL variable simultaneously with optical and near-infrared photometry, as well as spectroscopically through a complete modulation cycle, to definitely establish the change of the behaviour of the different layers of RRL star atmospheres during the Blazhko cycle.

\section*{Acknowledgments}
The comments of the anonymous referee helped to improve the lucidity and readability of the paper significantly. We are grateful to J. Nuspl and  L.G. Bal\'azs for fruitful discussions and A. Sollima for commenting the possible status of V48. GH acknowledges support by the Chilean Ministry for Economy, Development, and Tourism's Programa Iniciativa Cient\'ifica Milenio through grant IC~120009, awarded to the Millennium Institute of Astrophysics (MAS); by Proyecto Basal PFB-06/2007; by FONDECYT Regular 1141141, and by CONICYT-PCHA/Doctorado Nacional grant 2014-63140099.

\end{document}